\documentclass[12pt]{article}
\usepackage{lineno}

\usepackage[margin=1in]{geometry}
\usepackage{amssymb,amsmath,latexsym}                        
\usepackage{graphicx}
\usepackage{cite}
\usepackage{xcolor}
\usepackage{float}

\usepackage{amsthm}
\usepackage{multicol}
\usepackage{authblk}

\usepackage[colorlinks]{hyperref}

\usepackage{adjustbox}
\usepackage{bbm}
\usepackage{mathtools}

\usepackage{graphicx}
\usepackage{subcaption}
\date{} 

\begin{document}
\title{The study of a new time reconstruction method for MRPC read out by waveform digitizer}
\author[1,2]{Fuyue Wang}
\author[1,2]{Dong Han}
\author[1,2]{Yi Wang\footnote{Corresponding author. Email: yiwang@mail.tsinghua.edu.cn.}}
\author[1,2]{Yancheng Yu}
\author[1,2]{Pengfei Lyu}
\author[1,2]{Baohong Guo}
\affil[1]{\normalsize\it Department of Engineering Physics, Tsinghua University, Beijing 100084, China}
\affil[2]{\normalsize\it Key Laboratory of Particle and Radiation Imaging(Tsinghua University), Ministry of Education, Beijing 100084, China}

\maketitle

\begin{abstract}
The measurement of the $K^{\pm}$ production in the Semi-Inclusive Deep Inelastic Scattering (SIDIS) can provide further knowledge about the structure of nucleon, and thus it is purposed in the Solenoidal Large Intensity Device(SoLID) at Jefferson Lab(JLab). In this experiment, the identification of the kaons is planed to be accomplished with the Multi-gap Resistive Plate Chambers(MRPC), and the requirement for the time resolution is around 20 $\rm ps$. This is very challenging for the present MRPC systems (typical resolution 60 $\rm ps$), while in this paper, it is proved that the performance can be improved largely if the signal waveform is obtained and analyzed with a neural network method. In a cosmic ray experiment, the time resolution of a 6-gap 0.25$\rm mm$-thick MRPC reaches 36 $\rm ps$ with this method, and a even better performance is expected with a thinner MRPC.
\end{abstract}

\section{Introduction}
\label{sec:intro}
Understanding the strong interactions is an important topic in physics studies, but even with the powerful Quantum chromodynamics(QCD) theory, there is still a lot unknown about the nucleon structure. Among all the experiments, the Solenoidal Large Intensity Device(SoLID) at Jefferson Lab(JLab) shows great interests on the Semi-Inclusive Deep Inelastic Scattering(SIDIS), which is one of the main processes to extract the three-dimensional descriptions of partonic structures of the nucleon. A new measurement of the production of $K^{\pm}$ in the SIDIS process is proposed and the data will be combined with the existing $\pi^{\pm}$ measurements\cite{solid2014} to obtain valuable results.

The identification of the $K$ is planed to be accomplished with the Time-of-Flight(ToF) system which consists of the Multi-gap Resistive Plate Chambers(MRPC). In order to get a 3 $\sigma$ separation between pions and kaons up to momentum of 7 $\rm GeV/c$, the MRPCs in the experiments should have a time resolution of around 20 $\rm ps$. MRPC is a gas detector and has been widely used in many large physics experiments as the timing detector. The typical time resolution is around 60 $\rm ps$. The time uncertainty of the MRPC detector mainly comes from 3 sources: first is the intrinsic detector physics, which includes the ionization statistics and the uncertainty of the avalanche multiplication inside all the gas gaps. Second is the front-end electronics and other components. For present MRPCs, the signal is read out using the Time-Over-Threshold(ToT) method, which only reads out the threshold crossing time $t_c$ and the time interval the signal stays above that threshold $t_{tot}$. So the uncertainty comes from the variance of the time measurement and electronics noise of the front-end electronics(FEE) and the cables. Finally, since $t_c$ and $t_{tot}$ are given in TDC channels, the channel uncertainty ($20\sim 30 \rm ps$\cite{akindinov2004latest}) is another important source.

To improve the time resolution, we focus on using a fast amplifier and a waveform digitizer as the read out system, so an entire signal waveform is recorded instead of only the $t_c$ and $t_{tot}$. A new analysis method based on a neural network and machine learning algorithms is proposed and implemented to make the best use of the waveform. This method is an end-to-end solution of reconstructing the first interaction time between the incident particles and the working gas. A preliminary study of the neural network has proved that it is more robust with the noise and can achieve a better performance\cite{wang2018neural}. Therefore, two sets of the neural networks are studied and compared in this paper. The MRPC used has 6 gaps and each of them is 0.25 mm thick. The time resolution achieved with the simulation data is 20 ps, and that for the cosmic ray experiments is 36 ps, which is much better than the result obtained with the 0.25 mm-thick MRPC.

The paper is organized as follows: Sec.\ref{sec:NN} describes the method of the two neural networks. Sec.\ref{sec:simu} presents the experiment of the simulation conducted for the MRPC studied in this work. Sec.\ref{sec:result} shows the results and compares the different algorithms. Sec.\ref{sec:concl} concludes the paper.

\section{Method based on the neural network}
\label{sec:NN}
In recent years, the artificial neural network(NN) has shown its great power on solving highly non-linear problems such as pattern recognition, function approximation and prediction in many fields including the high energy physics analysis\cite{LSTMMagnetic,Atlasneural,Wang20181}. With the waveform digitizer, more information in the MRPC signal is extracted from each event, and in order to utilize this advantage, the method based on neural networks is proposed. It takes the input of all the points along the leading edge, learns the patterns of the signal and estimates the first interaction time inside the detector. Models of the networks are formed by training the labeled data and they are saved for estimation of the testing data. In this paper, we study two kinds of the network: the multilayer perceptron(MLP) and the Long Short Term Memory network(LSTM). 

\begin{figure}
    \centering
    \begin{subfigure}[b]{0.45\textwidth}
        \includegraphics[width=\textwidth]{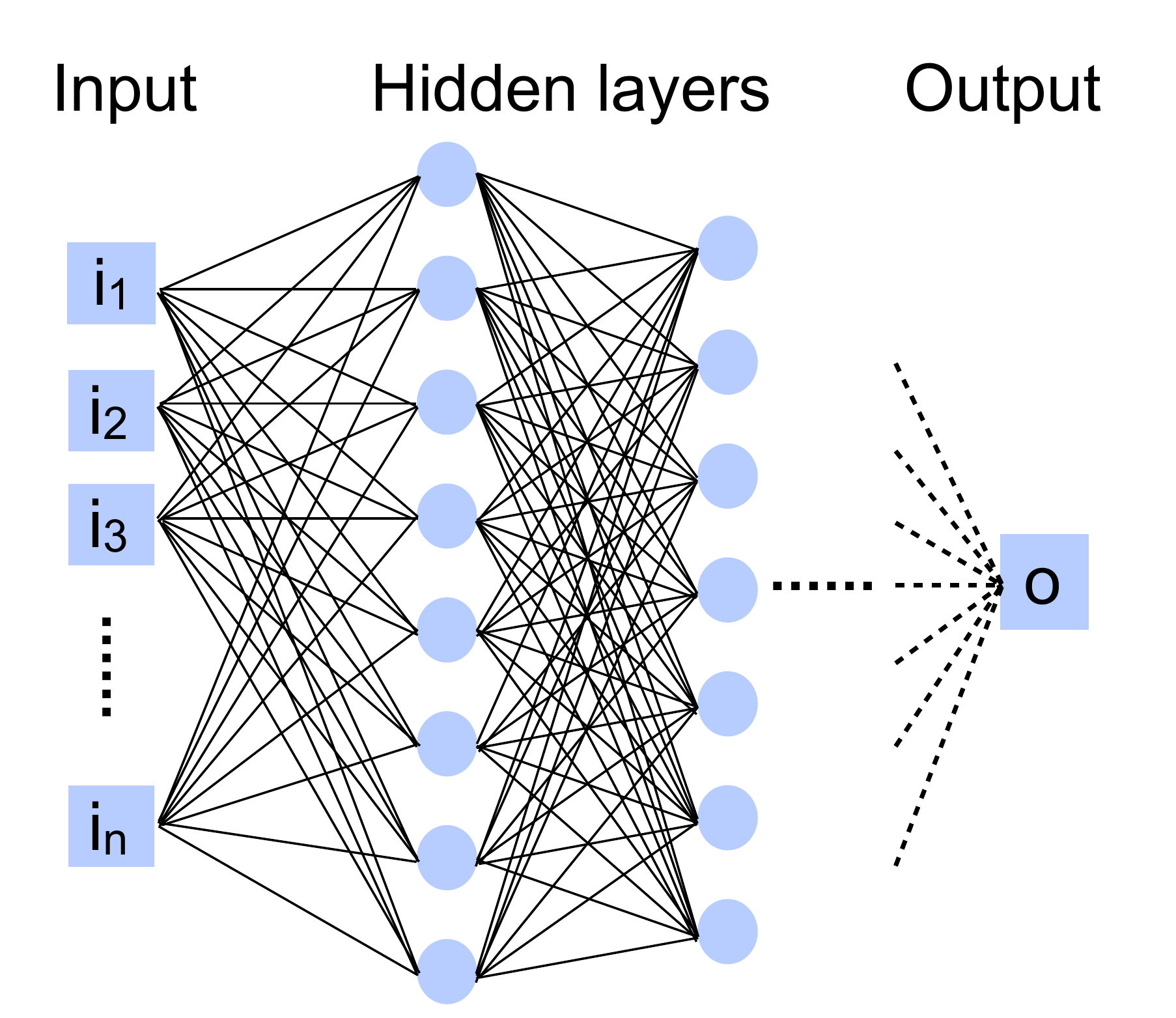}
        \caption{Multilayer perceptron}
        \label{fig:dnnstru}
    \end{subfigure}
    \begin{subfigure}[b]{0.45\textwidth}
        \includegraphics[width=\textwidth]{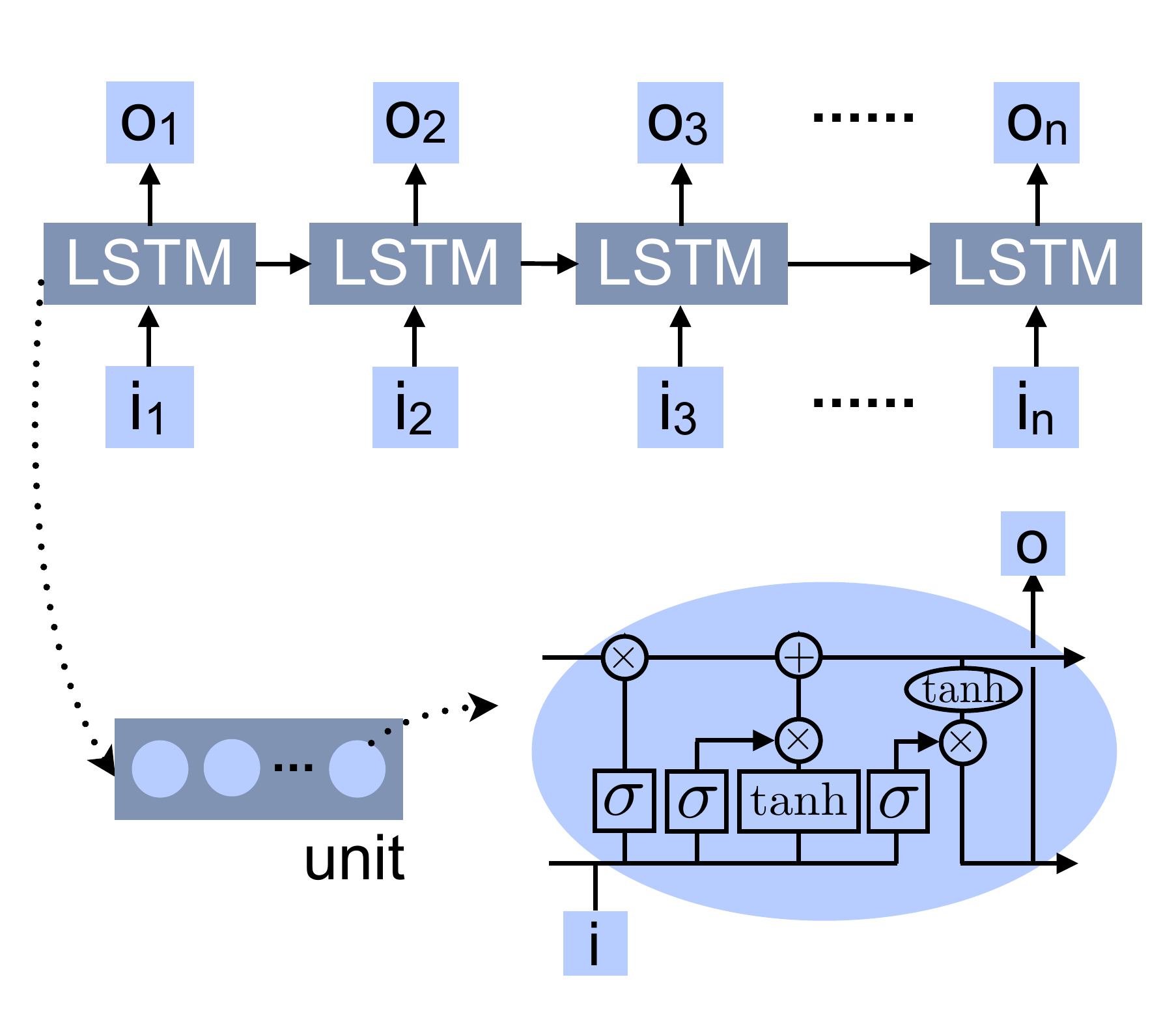}
        \caption{Long Short Term Memory network}
        \label{fig:lstmstru}
    \end{subfigure}
    \caption{The structure of two neural networks.}
    \label{fig:netstru}
\end{figure}

Fig.\ref{fig:netstru} shows the structure of two networks. The MLP is a feed-forward network and it consists of a great many simple interconnected neurons or nodes. These nodes are divided into several hidden layers and are connected with weights and bias terms. It has been shown that the MLP can approximately model any smooth, measurable relations between the input and output vectors\cite{mlphornik1989}. The training of the MLP works in a supervised manner, batches of the labeled data is presented to the network, and the weights and bias are adjusted until the model perfectly maps the inputs(points along the leading edge) and outputs(first interaction time). The MLP used in this study has 6 layers and each layer has about 16 points. The simplicity of the network architecture makes MLP converge very fast and easy to train. However, this may also limit its generalization ability, and thus the performance. Therefore, a feedback network----LSTM is also studied.

LSTM is one of the recurrent neural networks(RNN) and is often applied to process sequential data. Fig.\ref{fig:lstmstru} shows the unrolled structure of LSTM. $i_{1,2...n}$ are the inputs of the network which refer to the n points along the leading edge, while $o_{1,2...n}$ are the output at each time step. In the network considered in this study, the output of the last step $o_n$ is then fed into a 2-layer MLP, and the output of the last layer is regarded as the estimation of the first interaction time. The "LSTM" blocks include many basic units that describe the hidden cell states and the connecting weights. There are 4 "gates" in every unit. The first is the forget gate(f), which decides what information in the cell states is out of date and should be abandoned. The second is the input gate(I) which decides whether the input at this time step is valuable and should contribute to the cell state. The gate gate(G) decides how much of the input should be kept, and the output gate(O) decides how much of the cell state is relevant at current time step and then transmits the information to the output. The biggest difference between the feedback LSTM and the feed-forward MLP is that there is a path for the information to flow from the output of the previous steps to the present, and thus the network is dynamic. This kind of the structure is suitable for solving problems of the time related sequence data, and therefore it is adopted by our group to deal with the waveform of the MRPC signal.

\begin{figure*}[h!]
	\centering
	\includegraphics[width=0.5\textwidth]{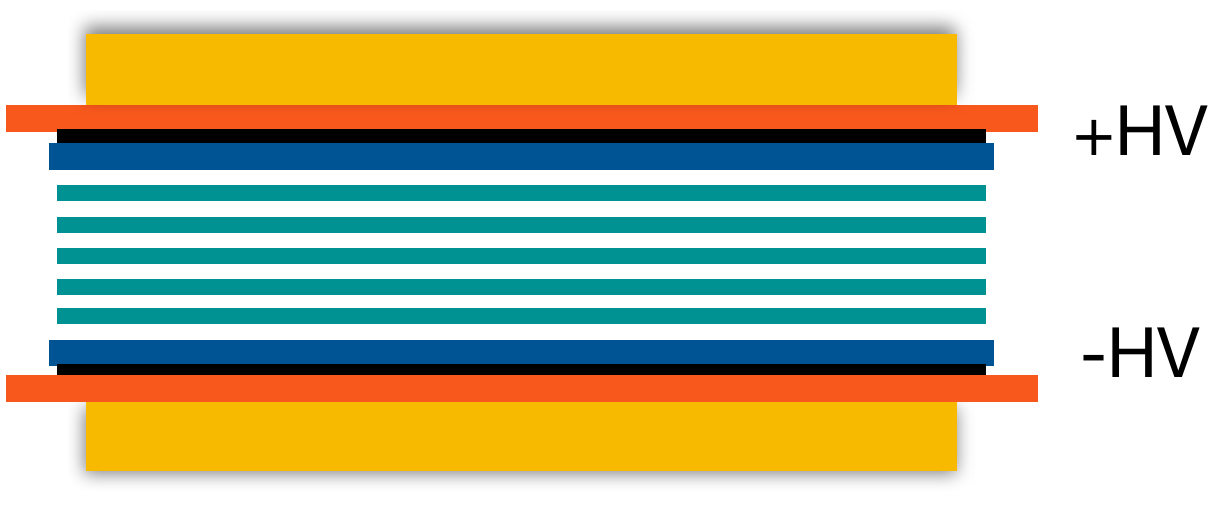}
	\caption{The structure of the 6-gap MRPC. The gap thickness is 0.25 $\rm mm$. }
	\label{fig:MRPCStru}
\end{figure*}

\section{The experiment and simulation}
\label{sec:simu}
The structure of the MRPC studied in this work is shown in Fig.\ref{fig:MRPCStru}. It has 1 stack and 6 uniform gas gaps. To be general, the gap is 0.25 mm thick, which is the most widely used thickness in large physics experiments. The gaps are separated by 5 0.7$\rm mm$-thick float glasses. The electrode glasses are on the top and bottom of the gas chamber and connect with the positive and negative high voltage respectively. Mylar films, PCB boards and the honeycomb panels are placed outside the two electrode glasses. The working gas of MRPC is 90\% $\rm C_2H_2F_4$, 5\% $\rm C_4H_{10}$ and 5\% $\rm SF_6$ at room temperature and under standard atmosphere.
\begin{figure*}[h!]
	\centering
	\includegraphics[width=0.5\textwidth]{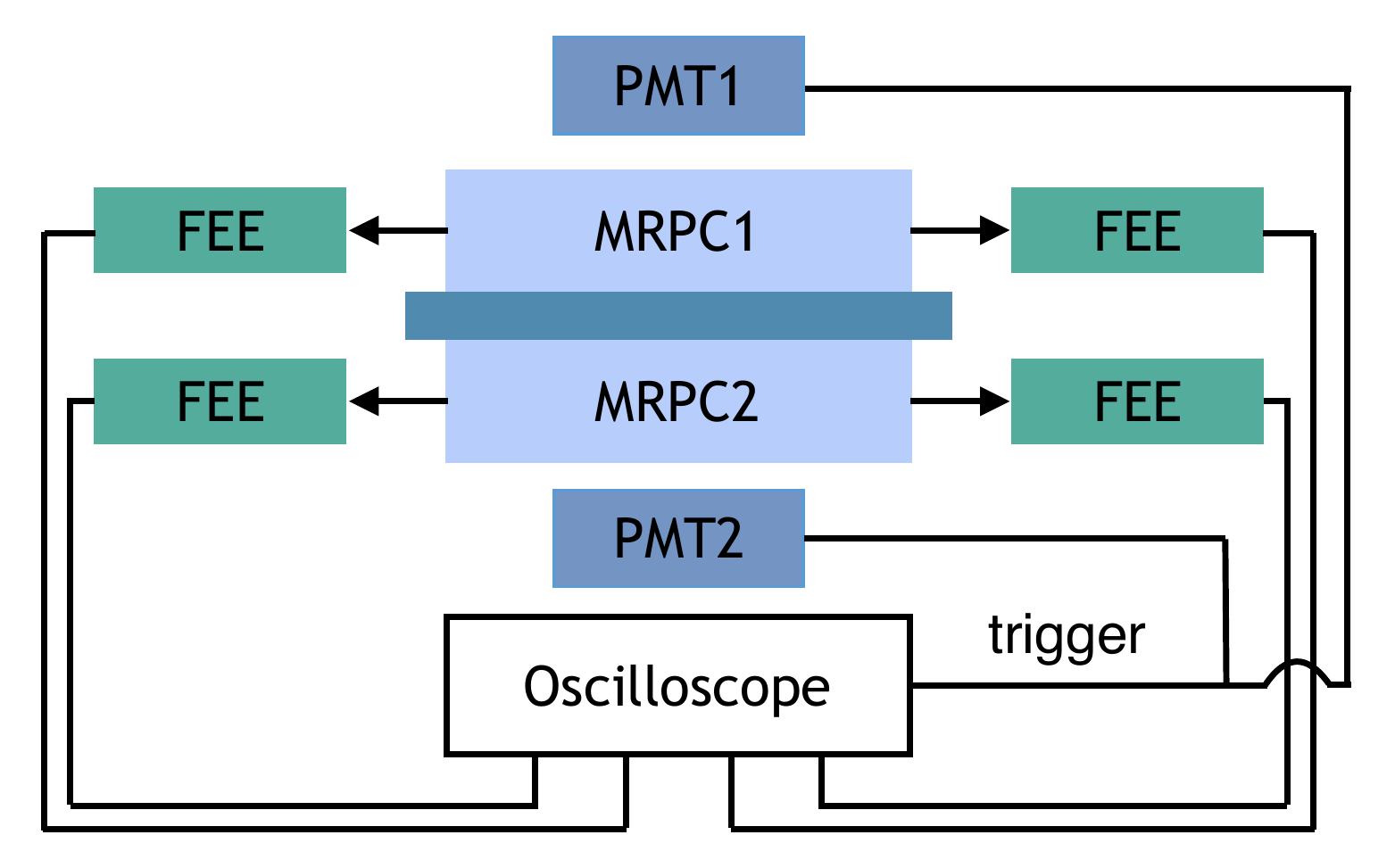}
	\caption{The experiment setup.}
	\label{fig:setup}
\end{figure*}

In the cosmic ray experiment conducted in our lab, two identical MRPCs are produced and placed one on top of the other. The setup of the experiment is shown in Fig.\ref{fig:setup}. MRPCs are vertically aligned and separated by a block. The signals of the detectors are read out from both ends of the strips and are collected by a fast amplifier made by Texas Instruments\cite{Tielectronics}. The amplified signal is read out and digitized by an oscilloscope which has a bandwidth of 1 GHz and sampling rate of 10 GS/s. The coincidence of 2 PMTs is used as the trigger. 

Data given by the oscilloscope is analyzed by the models of the two neural networks described in Sec.\ref{sec:NN}. The labeled data used to train the network comes from an accurate simulation of the signal waveform. The simulation in this work is based on a dedicated standalone framework built by our group\cite{SimulationFuyue}. The geometry of the simulated MRPC is exactly the same as the one in the experiment. Cosmic muons with a mean energy of 4 $\rm GeV$ is simulated and the angle of the particles is nearly perpendicular to the MRPC detector with a small variance. When the muon goes through the gas chamber, it interacts with the working gas and the electron-ion pairs are created. The ionized electrons drift and get multiplied under the electric field and induce a current signal on the read out strip. The simulation of the front-end electronics and the waveform digitizer is also included and takes the same parameters as the experiments.

\begin{figure*}[h!]
	\centering
	\includegraphics[width=0.5\textwidth]{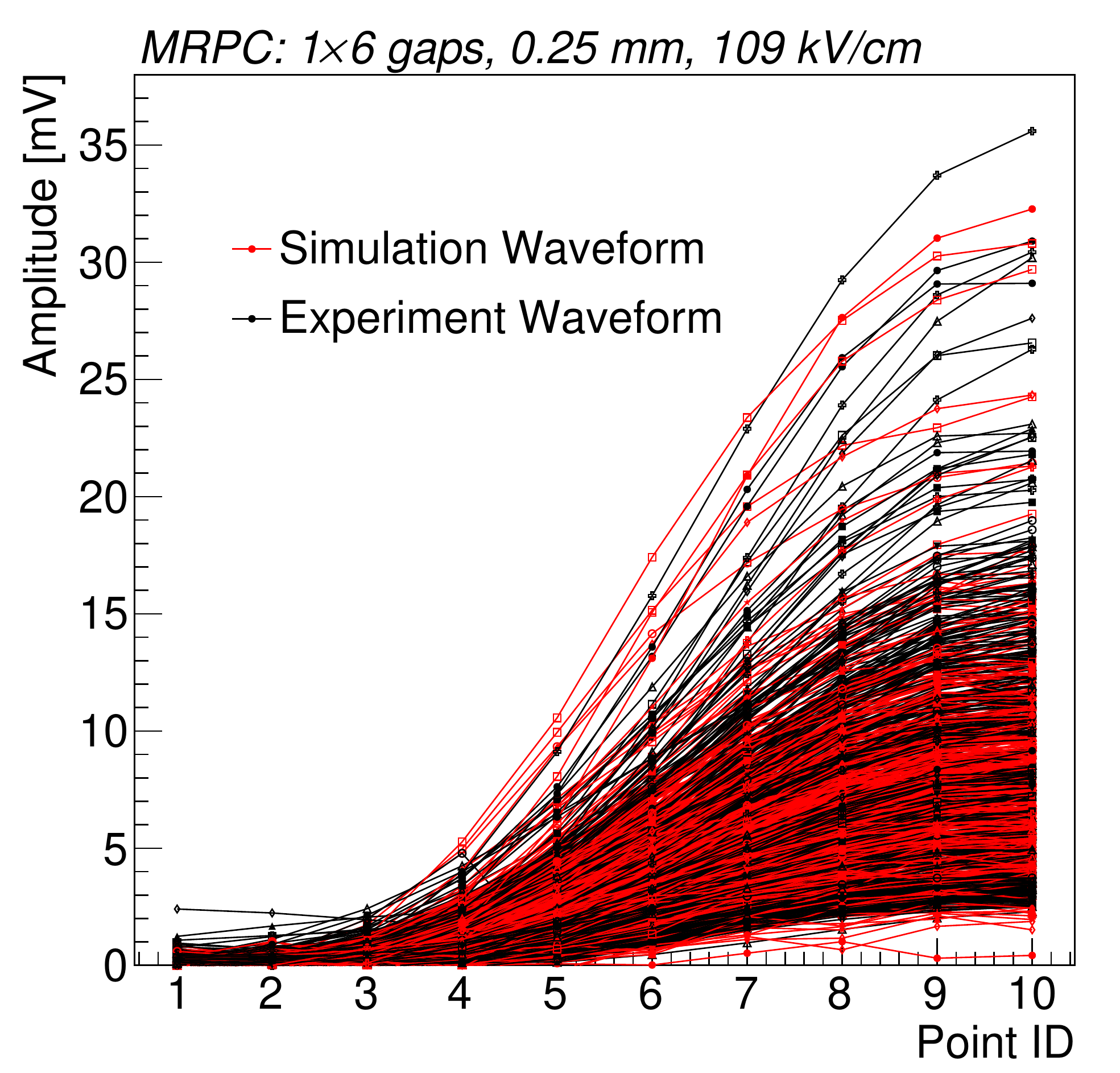}
	\caption{The waveform. The red markers and lines are from the simulation, while the black markers and lines are from the experiment. }
	\label{fig:MRPCSwaveform}
\end{figure*}

Fig.\ref{fig:MRPCSwaveform} shows the data given by the digitizer. The red markers and lines are from the simulation, while the black markers and lines are from the experiment. The length of the leading edge is around 700$\sim$800 ps, so there are about 7 points above the threshold along the edge. The shapes of the red and black curves in Fig.\ref{fig:MRPCSwaveform} agree well in general, which means the simulation data should be a good representative of the experiment. This indicates that the knowledge of the leading edge gained from the simulation data by the neural networks applies to the experiment, and can be used to make the estimation for the experiment. 

\section{Results}
\label{sec:result}
We trained the network with 7 points on signal's leading edge and estimate the length of the edge $t_l$, which is defined to be the time interval from the first interaction to the wave peak. Hence, by obtaining the peak time $t_p$ from the waveform, we can reconstruct the first interaction time as $t_a=t_p-t_l$. The training and testing are based on the Tensorflow\cite{abadi2016tensorflow}. Over 200,000 events are used as the training data, and 100,000 are for testing. It takes about 10 minutes for the MLP to converge and 30 minutes for the LSTM on a GPU. Dropout is applied to both the neural networks as the regularization to prevent overfitting. 

\begin{figure}
    \centering
    \begin{subfigure}[b]{0.45\textwidth}
        \includegraphics[width=\textwidth]{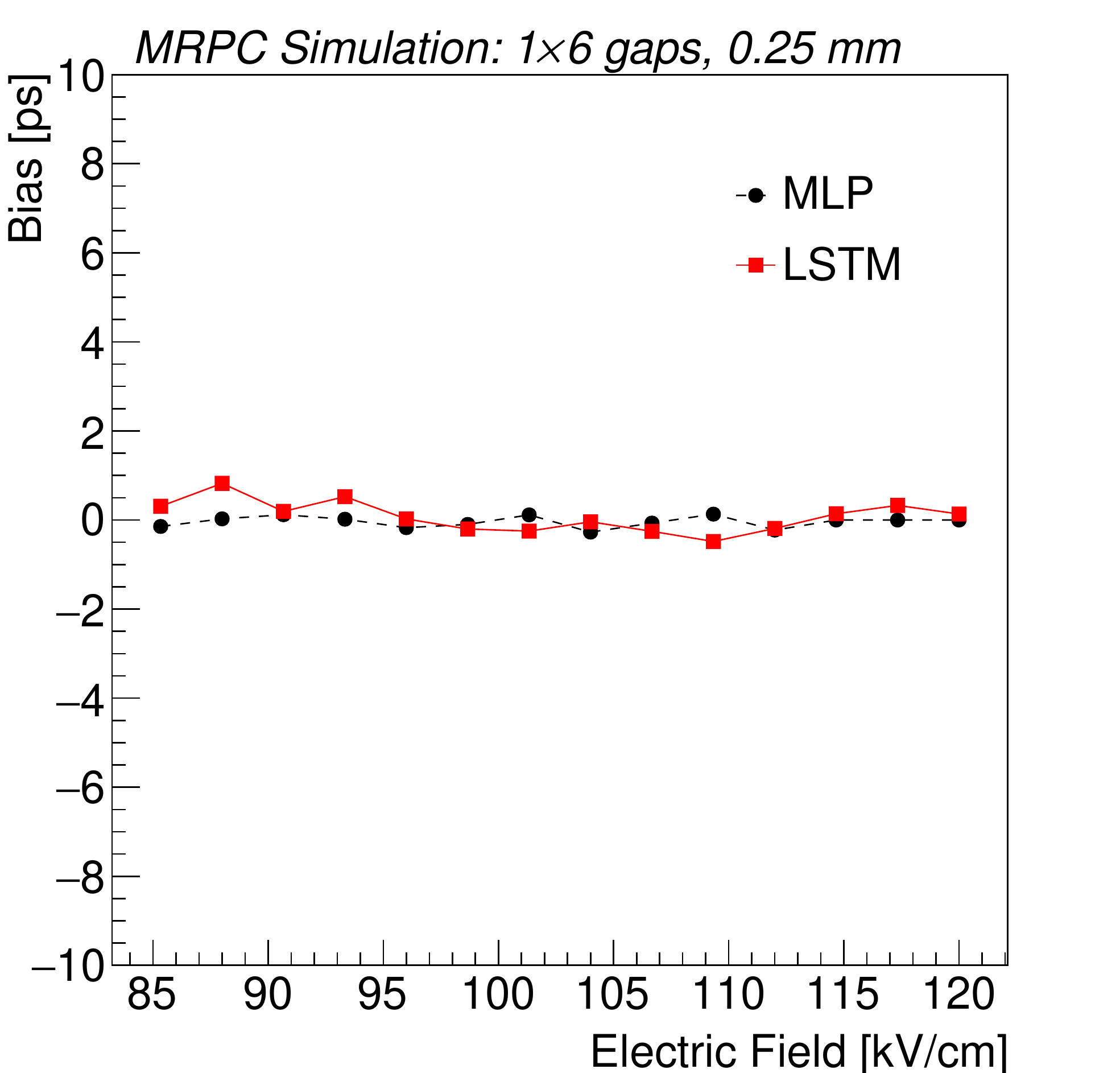}
        \caption{Bias of the estimators.}
        \label{fig:bias}
    \end{subfigure}
    \begin{subfigure}[b]{0.45\textwidth}
        \includegraphics[width=\textwidth]{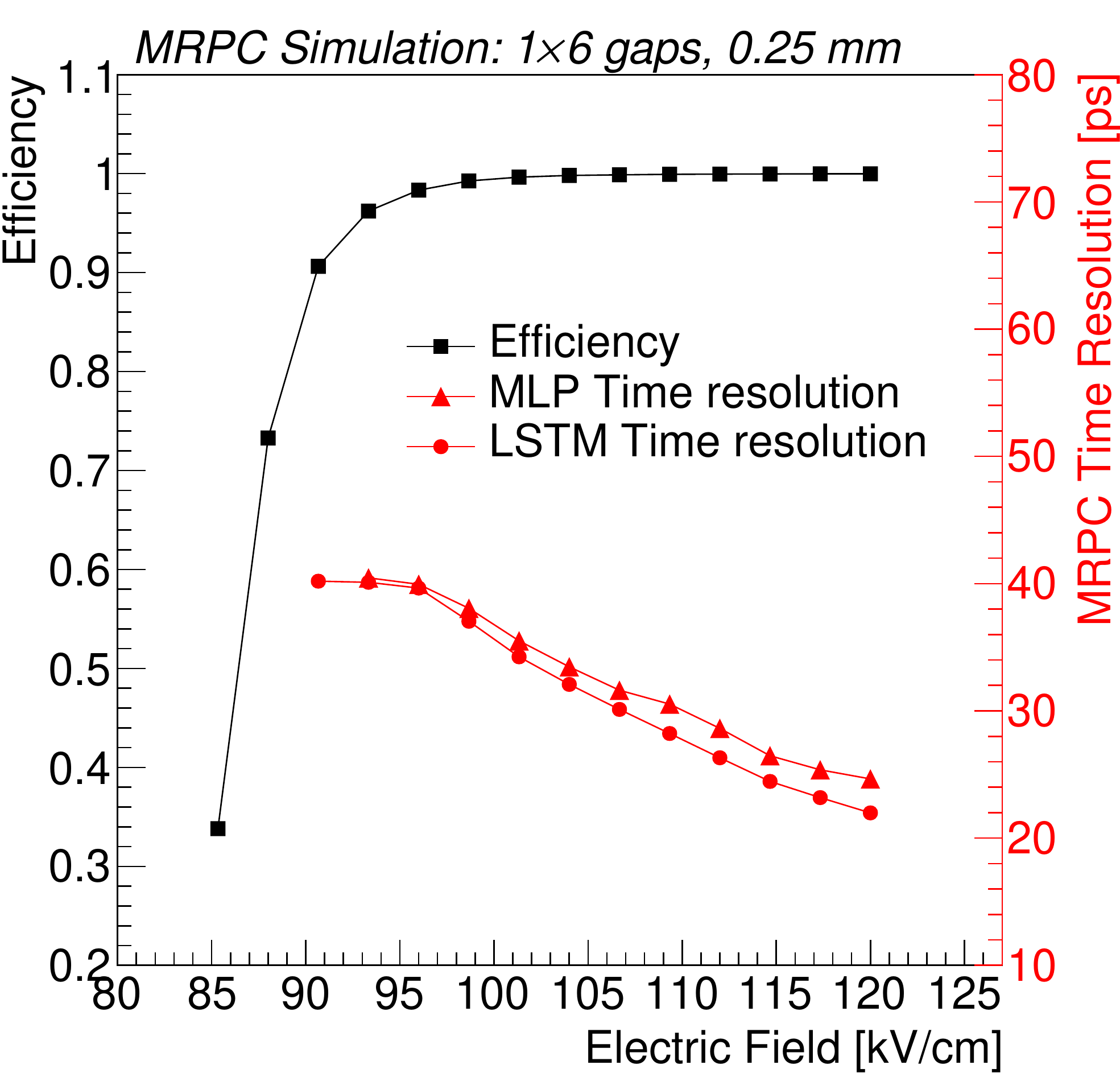}
        \caption{Efficiency and the time resolution.}
        \label{fig:reso}
    \end{subfigure}
    \caption{Accuracy and precision of two neural networks.  }
    \label{fig:performance}
\end{figure}

\begin{figure*}[h!]
	\centering
	\includegraphics[width=0.5\textwidth]{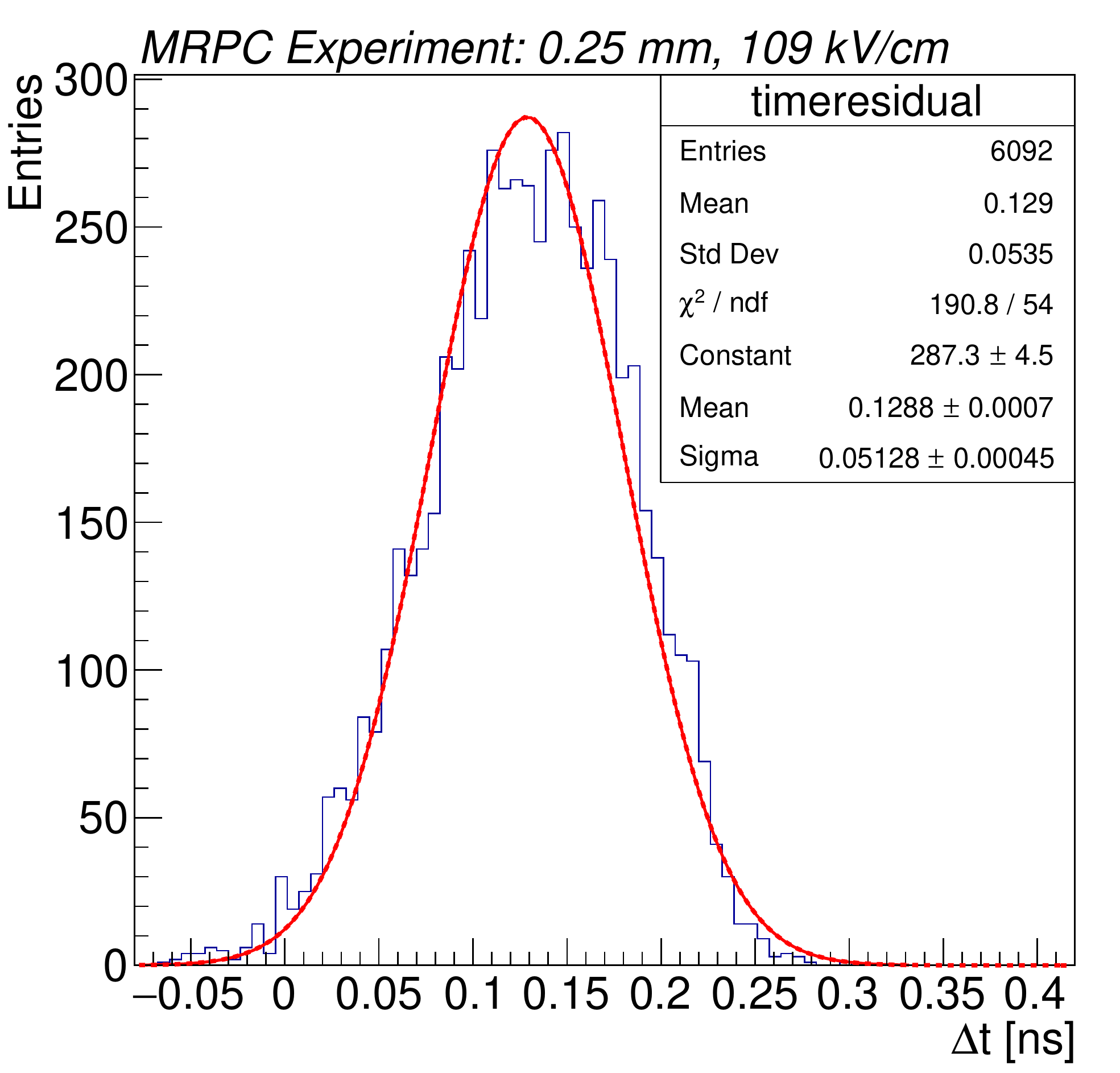}
	\caption{The result of the LSTM}
	\label{fig:lstm}
\end{figure*}

\begin{figure}
    \centering
    \begin{subfigure}[b]{0.45\textwidth}
        \includegraphics[width=\textwidth]{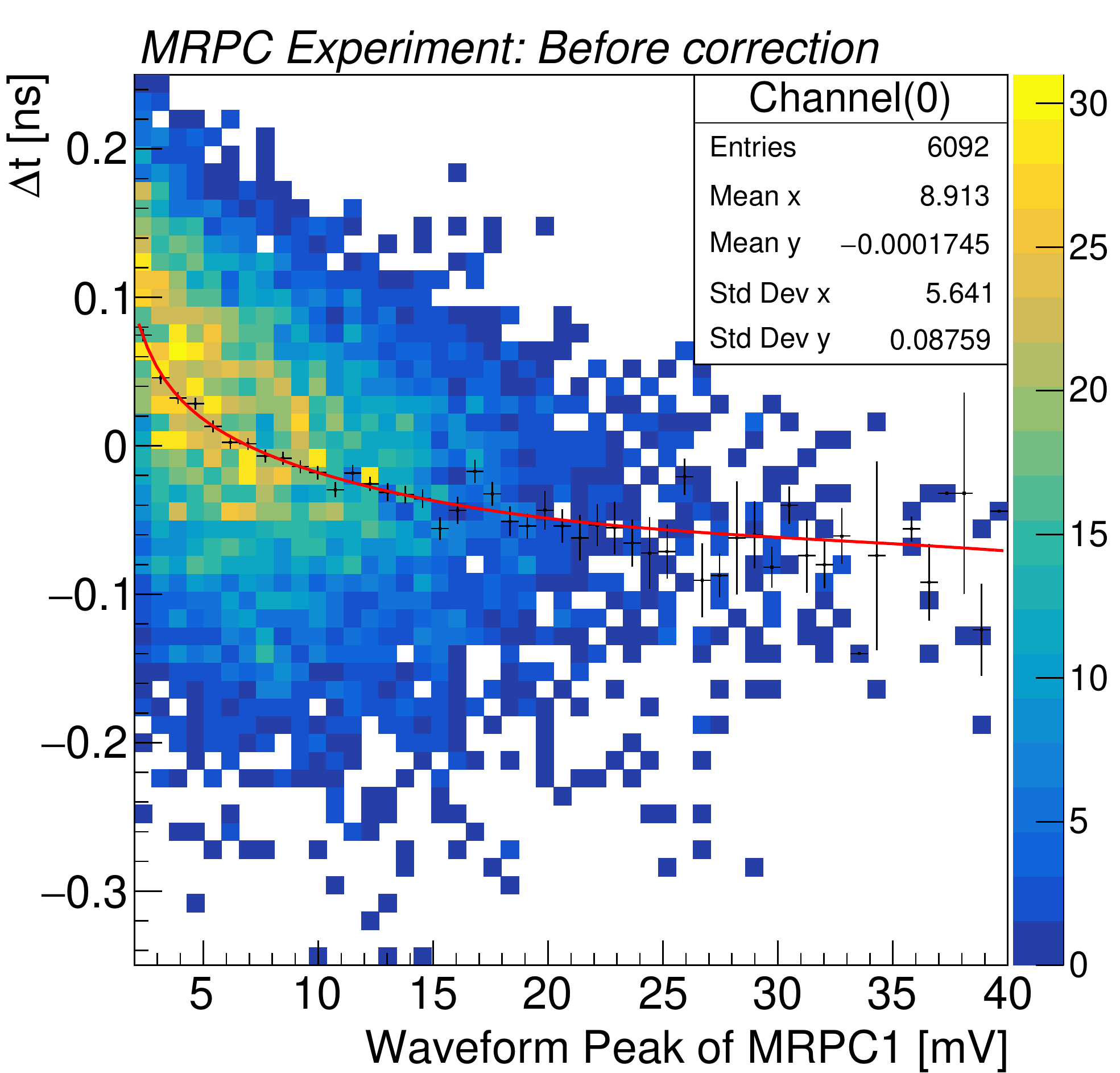}
        \caption{}
        \label{fig:slewa}
    \end{subfigure}
    \begin{subfigure}[b]{0.45\textwidth}
        \includegraphics[width=\textwidth]{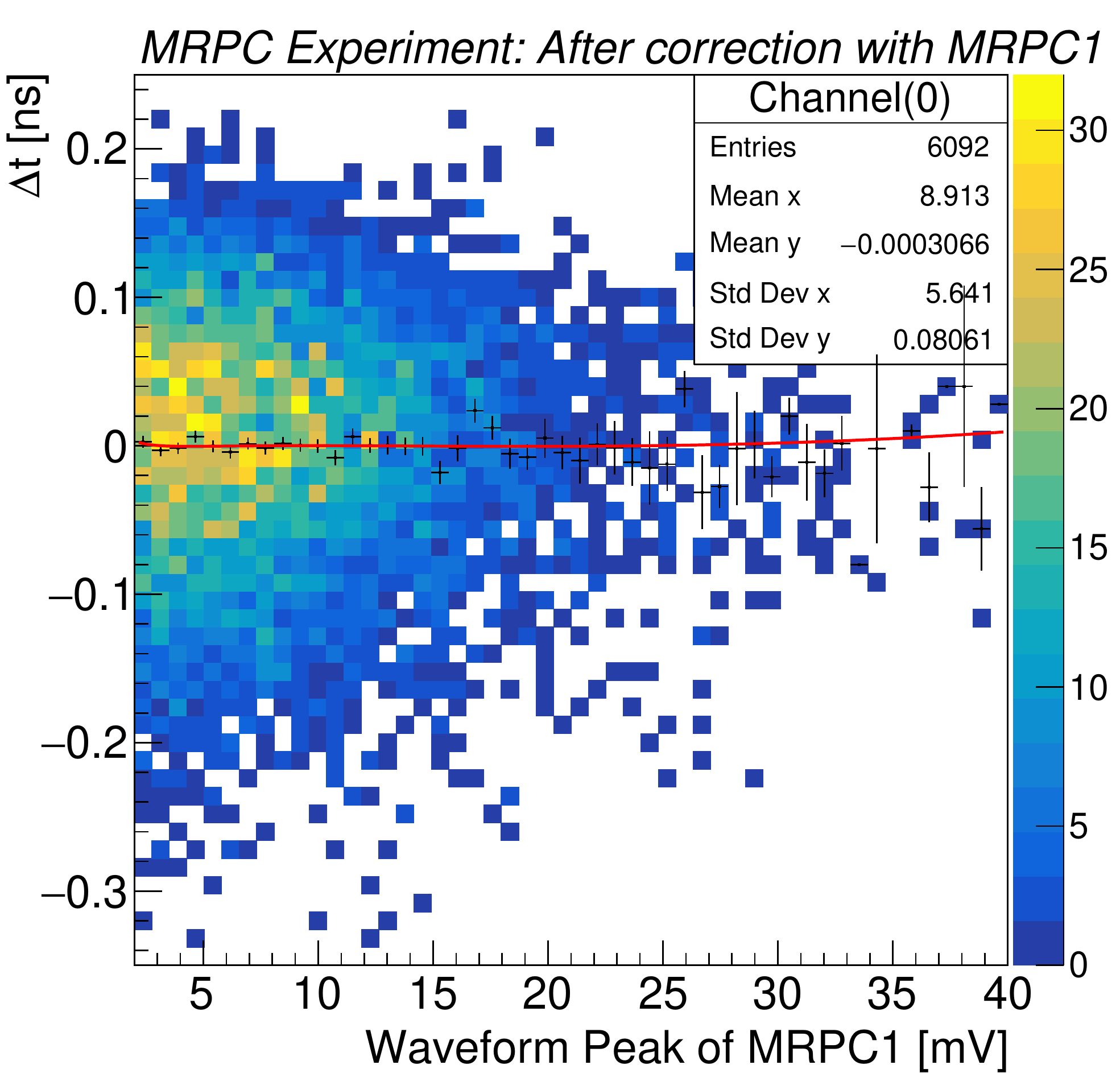}
        \caption{}
        \label{fig:slewb}
    \end{subfigure}
    \begin{subfigure}[b]{0.45\textwidth}
        \includegraphics[width=\textwidth]{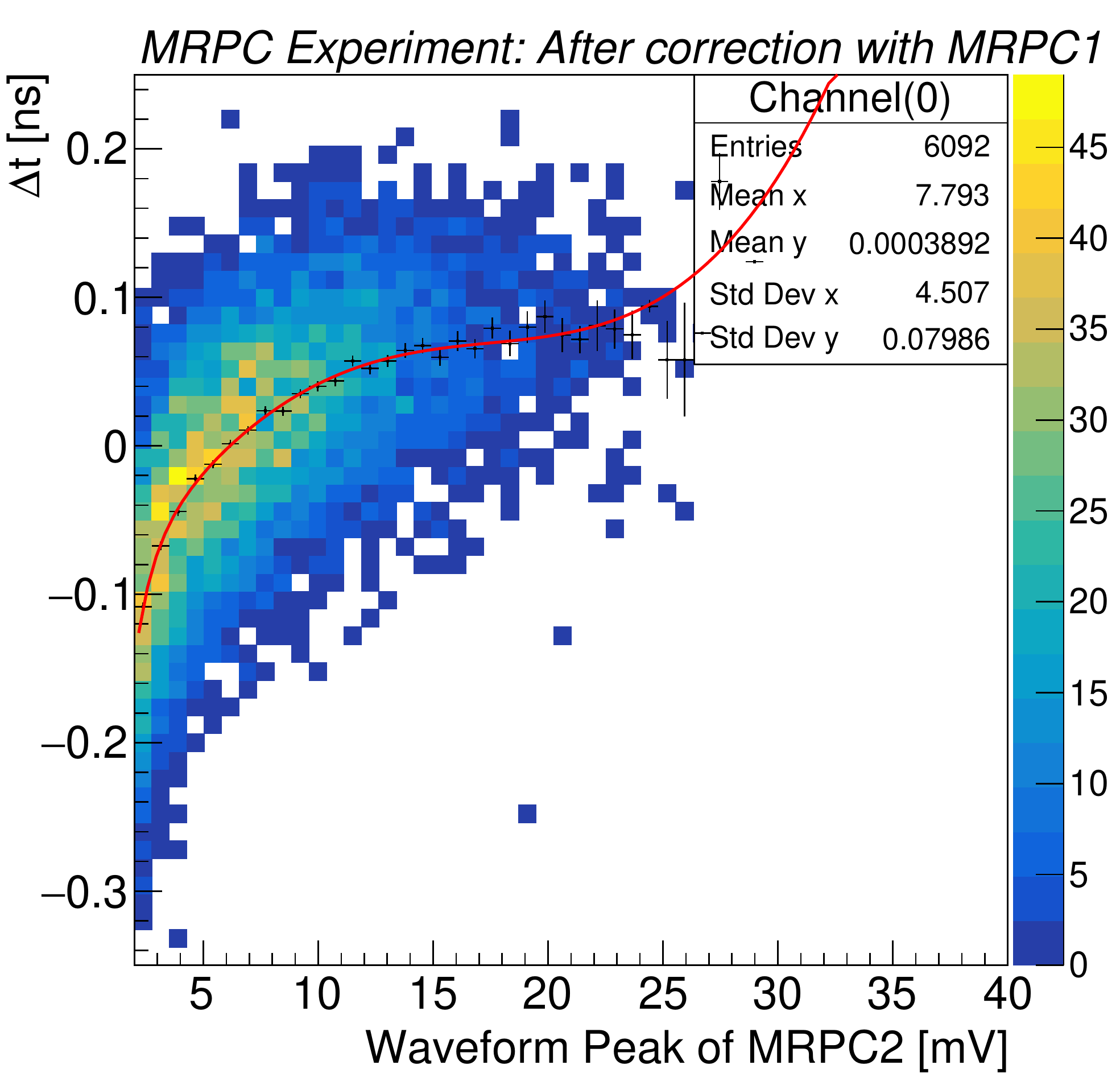}
        \caption{}
        \label{fig:slewc}
    \end{subfigure}
    \begin{subfigure}[b]{0.45\textwidth}
        \includegraphics[width=\textwidth]{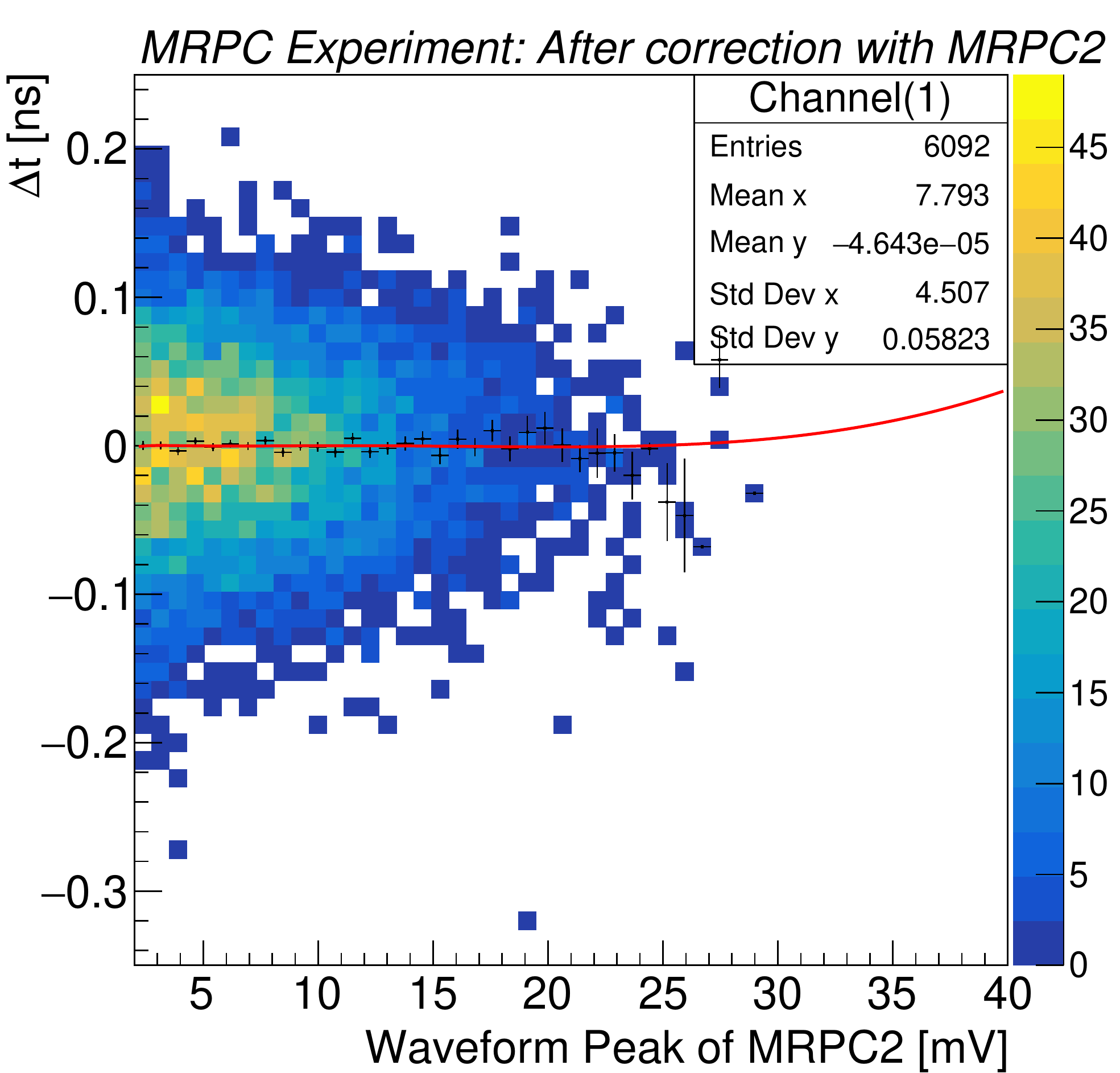}
        \caption{}
        \label{fig:slewd}
    \end{subfigure}
    \caption{The slewing correction for ToT method. (a) shows the 2D distribution of $\Delta t$ and the wave peak of MRPC1, while (b) shows this distribution after the correction with MRPC1. (c),(d) shows the 2D distribution of $\Delta t$ and the wave peak of MRPC2 before and after the correction with MRPC2.}
    \label{fig:slewing}
\end{figure}

Accuracy and precision are two of the most important indicators of the performance for any detectors. The result of the two obtained purely from the simulation are shown in Fig. \ref{fig:performance}. The bias shown on the Y axis of Fig.\ref{fig:bias} is defined to be:
\begin{equation}
\label{eq:bias}
bias=t_{estimate}-t_{true}
\end{equation}
which indicates how accurate the detectors can be with the network. The biases of both the networks are below 0.5 ps with respect to the electric field in the gas gap, which manifests a good accuracy of the neural network based method. The precision or the time resolution achieved with the networks is shown in Fig.\ref{fig:reso} with the red markers and lines. The time resolution gets improved as the electric field $E$ becomes higher. The results of the two networks at low $E$ are similar, but differences appear with high electric field. As expected, LSTM has a lower time resolution than MLP, and the best resolution we get from a 0.25 $\rm mm$ thick, 6-gap MRPC is 22 $\rm ps$. The resolution shown in Fig.\ref{fig:reso} is achieved by cutting off those extremely small signals, which causes a drop of the detector efficiency. The black line and dots in Fig. \ref{fig:reso} shows the efficiency. 

The simulation stops around $E=120$ kV/cm, because in the experiments, chances are great for streamers to happen under the electric field higher than this. The MRPCs are expected to work in avalanche mode, so studies of streamers  is beyond the scope of this work.

As for the experiment data, we only shows the result obtained by the LSTM, since it is better than the MLP. Fig.\ref{fig:lstm} shows the distribution of $\Delta t$, which is defined to be the difference of the estimated first interaction time of two MRPCs. The time difference is calculated in order to eliminate the uncertainty of the trigger, and since two MRPCs are identical, the time resolution should be:
\begin{equation}
\label{eq:bias}
\sigma_{MRPC}^{LSTM}=\sigma(\Delta t)/\sqrt{2}=51.28/\sqrt{2}=36.26\;\rm ps
\end{equation}

The same experiment data is also analyzed with the ToT method, which extracts the $t_c$ and the wave peak under a fixed threshold of around 1.5 mV from a fitted waveform. The difference of the wave peak for different signals generates a time walk to $t_c$, and thus should be corrected by the slewing correction. For the test system in this work, $\Delta t$ should be corrected by both the wave peak of MRPC1 and MRPC2. The correction is shown in Fig.\ref{fig:slewing}. Fig.\ref{fig:slewa} shows the 2D distribution of $\Delta t$ and the wave peak of MRPC1, while Fig.\ref{fig:slewb} shows this distribution after the correction with MRPC1. The distribution of $\Delta t$ and the wave peak of MRPC2 before and after the correction with MRPC2 is shown in Fig.\ref{fig:slewc} and Fig.\ref{fig:slewd}. This 2-step correction is iterated several times until the time walk caused by the pulse height is eliminated. The final resolution achieved with the this method is:
\begin{equation}
\label{eq:bias}
\sigma_{MRPC}^{ToT}=\sigma(\Delta t)/\sqrt{2}=58.23/\sqrt{2}=41.17\;\rm ps
\end{equation}

Moreover, Fig.\ref{fig:lstm} shows that the mean value of $\Delta t^{LSTM}=129\;\rm ps$ and that of the ToT method is $\Delta t^{ToT}=131.5\;\rm ps$. The value of $\Delta t$ should be related to the distance between the two MRPCs, and the results obtained from the method is consistent with the experiment. This results prove that the neural network based method gives an accurate estimation of the first interaction time inside the detector, and the time resolution is also better than the ToT, which is the state-of-the-art method. The MRPC studied in this work is 0.25 mm thick, and it is expected that the time resolution of a thin detector, like 0.1 mm thick, should be even better. 

\section{Conclusions}
\label{sec:concl}
A new time reconstruction method based on the neural networks is proposed for the MRPC detector. The network learns the pattern of the data from the simulation and applies the knowledge to the experiment in order to estimate the particles' first interaction time. Two kinds of the networks MLP and LSTM are studied and compared. LSTM, the better network, is used in the experiment and gives an accurate and precise result for the detector. The best time resolution of this 0.25 mm thick MRPC is 36 ps, and can be improved with a thinner detector. 

\section{Acknowledgments}
The work is supported by National Natural Science Foundation of China under Grant No.11420101004, 11461141011, 11275108, 11735009. This work is also supported by the Ministry of Science and Technology under Grant No. 2015CB856905, 2016 YFA0400100.

\bibliographystyle{elsarticle-num}
\bibliography{myrefs.bib}{}

\end{document}